\documentclass[10pt]{article}
\usepackage[utf8]{inputenc}
\usepackage[T1]{fontenc}
\usepackage{amsmath}
\usepackage{amsfonts}
\usepackage{amssymb}
\usepackage[version=4]{mhchem}
\usepackage{stmaryrd}
\usepackage{graphicx}
\usepackage[export]{adjustbox}
\usepackage{csquotes}
\usepackage{hyperref}
\usepackage{authblk}
\usepackage{color}
\usepackage{amsthm}
\usepackage{orcidlink}

\newcommand{\HRule}[1]{\rule{\linewidth}{#1}}

\title{ \normalsize \textsc{}
    \HRule{2pt} \\
    \LARGE \textbf{Closed timelike curves in \\
    $\mathcal{PT}$-symmetric wormholes} 
    \HRule{2.0pt}}

\author{\small Hicham Zejli\,\orcidlink{0009-0006-8886-7101}}
\affil{\itshape Independent Researcher, France}

\date{}

\begin{document}

\maketitle

\begin{abstract}
We investigate a modified Einstein-Rosen wormhole model, made unidirectionally traversable through a bimetric geometry defined by two regular metrics, $g^{(+)}$ and $g^{(-)}$, and characterized by $\mathcal{PT}$ symmetry combining time reversal ($t \to -t$) and spatial inversion ($\vec{x} \to -\vec{x}$). In this framework, two distinct spacetime regions are identified at the wormhole throat ($r = \alpha$) via $\mathcal{PT}$ symmetry, forming a single spacetime sheet. This model employs Eddington-Finkelstein coordinates to eliminate coordinate singularities at the throat, enabling traversability with a lightlike membrane of exotic matter at the junction to satisfy the Einstein field equations, similar to other traversable wormhole models. We extend this model by coupling two such wormholes to generate closed timelike curves (CTCs), made possible by the opposing causal orientations defined by the two metrics, while adhering to Novikov’s self-consistency principle. An effective theory is developed for a scalar field crossing the wormhole, yielding $\mathcal{PT}$-symmetric Klein-Gordon equations with a real energy spectrum ensured by pseudo-unitarity, consistent with quantum mechanical dynamics. These results open new avenues for exploring the effects of $\mathcal{PT}$ symmetry on causality and the quantization of scalar fields in traversable geometries.\\
\end{abstract}


\section{Introduction}

Einstein–Rosen bridges, also known as ``Schwarzschild wormholes'', were first introduced by Einstein and Rosen in 1935 as a non-singular representation of particles, modeling a physical space with two identical sheets connected by a bridge~\cite{einstein1935er}. The original Einstein–Rosen wormhole corresponds to the maximally extended Schwarzschild metric (Kruskal–Szekeres extension), where the ``throat'' region connects two asymptotically flat universes. However, this original bridge is not traversable: in Schwarzschild coordinates, a traveler cannot reach the throat radius $r = \alpha$ in finite coordinate time $t$, as crossing the horizon would require an infinite coordinate time~\cite{koiran2021,Oppenheimer1939}. Furthermore, the identification of the two sheets posed a ``gluing problem'': directly attaching the two spacetime halves effectively reduced the space to a single sheet, making the second sheet redundant~\cite{koiran2021}.\\

Wormholes have been extensively studied in the literature as a potential solution to these challenges, notably in the work of Guendelman~\cite{guendelman2010wormholes}, who explored the concepts of wormholes and child universes within the framework of classical gravity. He also revisited the Einstein–Rosen bridge by proposing formulations of wormholes with lightlike membranes~\cite{guendelman2016einstein}. \\

In this vein, a modified Einstein–Rosen bridge model was proposed to address these challenges~\cite{koiran2021,koiran2024}. Drawing on earlier work with Eddington coordinates~\cite{eddington1924}, this model introduces a modified metric incorporating a cross term $\mathrm{d}r\,\mathrm{d}t$, which eliminates the coordinate singularity at the throat. Specifically, instead of using the divergent Schwarzschild time, a linear combination of $t$ and $r$ is employed, allowing free fall through the throat in finite time for an external observer. This modification ensures that the bridge is traversable while behaving like a one-way membrane: traversal is allowed in only one direction. Moreover, geodesics emerge in finite time, resolving the classical issue of infinite ``escape time'' associated with Schwarzschild black holes~\cite{misner1973gravitation,Oppenheimer1939}. The modified metric, non-degenerate at the throat ($r = \alpha$), thus resolves the topological gluing problem that plagued the original Einstein–Rosen bridge~\cite{einstein1935er,koiran2021}\footnote{Or its traversable extensions~\cite{morris1988,Visser1995}}. While our model utilizes Eddington-Finkelstein coordinates to ensure geometric regularity, it requires a lightlike membrane of exotic matter at the throat to satisfy the Einstein field equations, as proposed in the modified Einstein-Rosen bridge model \cite{koiran2024} and supported by the analyses of Guendelman et al. \cite{guendelman2010, guendelman2016einstein}.

\begin{figure}[h]
\centering
\includegraphics[width=0.5\textwidth]{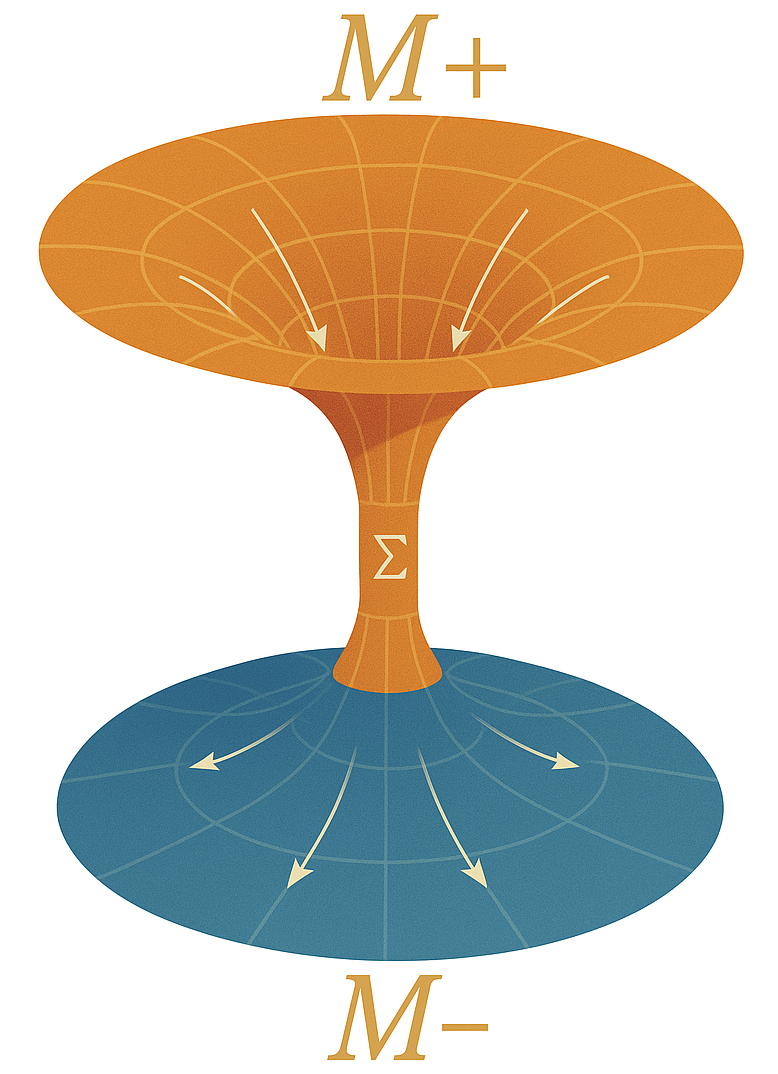}
\caption{Schematic representation of $\mathcal{PT}$ symmetry in the bimetric wormhole. The regions $\mathcal{M}_+$ and $\mathcal{M}_-$ are identified at the throat ($r = \alpha$) via $\mathcal{PT}$ symmetry, forming a single spacetime sheet~\cite{koiran2024}. A particle crossing the throat undergoes time ($t \to -t$) and spatial ($\vec{x} \to -\vec{x}$) inversions, potentially appearing in the past of its starting point if a second wormhole is introduced to form a temporal loop (see Figure~\ref{fig:CTC}).}
\label{fig:Trou_ver}
\end{figure}

A striking feature of this model is that, upon crossing the throat, a particle experiences a reversal of its temporal evolution from the perspective of an external observer. In other words, the solution exhibits $\mathcal{PT}$ symmetry: the metric of the outgoing region appears as the image of the incoming region under a combined time reversal ($\mathcal{T}$) and an appropriate spatial inversion ($\mathcal{P}$) at the throat, with the two regions, $\mathcal{M}_+$ and $\mathcal{M}_-$, identified to form a single spacetime sheet~\cite{koiran2024} (Figure~\ref{fig:Trou_ver}). In our extension, where two such wormholes are coupled\footnote{The mutual interaction between the two bridges is neglected in this idealized scenario.} (Figure~\ref{fig:CTC}), this behavior enables the formation of closed worldlines: a traveler crossing the wormholes could, in theory, return to the past of their starting point, opening the possibility of closed timelike curves (CTCs). Closed timelike curves are a well-established concept in general relativity: since Gödel’s discovery of a rotating universe containing such curves in 1949~\cite{godel1949}, it has been known that general relativity theoretically permits their existence, although their physical realization remains speculative. Traversable wormholes have been proposed as a mechanism for time travel; for instance, Morris, Thorne, and Yurtsever demonstrated that a wormhole can function as a time machine if a temporal offset exists between its ends~\cite{morris1988}.\\

In this paper, we revisit and enhance the analysis of the one-way $\mathcal{PT}$-symmetric wormhole model inspired by~\cite{koiran2024}. We systematically examine the following aspects:
\begin{itemize}
    \item The bimetric\footnote{In the context of~\cite{koiran2024}, ``bimetric'' refers to a single spacetime described by two metric forms, where the metric is continuous across the throat but its first derivatives exhibit a discontinuity, compensated by a lightlike membrane of exotic matter to satisfy the Einstein field equations~\cite{guendelman2010,guendelman2016einstein,koiran2024}.} $\mathcal{PT}$-symmetric geometric construction with two sheets, its one-way traversability, and its interpretation;
    \item The possibility of closed geodesics crossing the throat and returning to the past;
    \item Issues of causal consistency and Novikov’s self-consistency principle, which prevents paradoxes;
    \item The Einstein field equations and associated energy conditions for this modified bridge, including the role of a thin shell of exotic matter at the throat;
    \item The effective theory and quantization of a scalar field crossing the $\mathcal{PT}$-symmetric wormhole, yielding Klein-Gordon equations with pseudo-unitary dynamics, inspired by Bender and Kuntz.
\end{itemize}

Although speculative, this model could open avenues for testing CTCs through theoretical signatures, such as anomalies in gravitational waves or quantum fluctuations, providing a framework to explore the boundaries of causality in physics.\\

Each section includes a conceptual and physical review, drawing on recent and classical literature (Einstein–Rosen bridges, Morris–Thorne wormholes, $\mathcal{PT}$ symmetry in quantum physics, Novikov’s principle, Gödel’s universe, etc.), to ensure a rigorous and up-to-date presentation of the proposed model.

\section{Closed timelike loop in the $\mathcal{PT}$-Symmetric bimetric wormhole model}

As introduced earlier, the modified Einstein–Rosen bridge~\cite{koiran2024} is traversable in finite time due to the use of Eddington–Finkelstein coordinates~\cite{misner1973gravitation}, resolving the topological gluing problem of the original version proposed by Einstein and Rosen~\cite{einstein1935er}. This model, characterized by $\mathcal{PT}$ symmetry and a bimetric structure, forms the foundation of our analysis. This section summarizes its geometric properties and explores avenues for further investigation of its implications.

\subsection{Two $\mathcal{PT}$-symmetric bridges with sheet identification}

\begin{figure}[h]
\centering
\includegraphics[width=\textwidth]{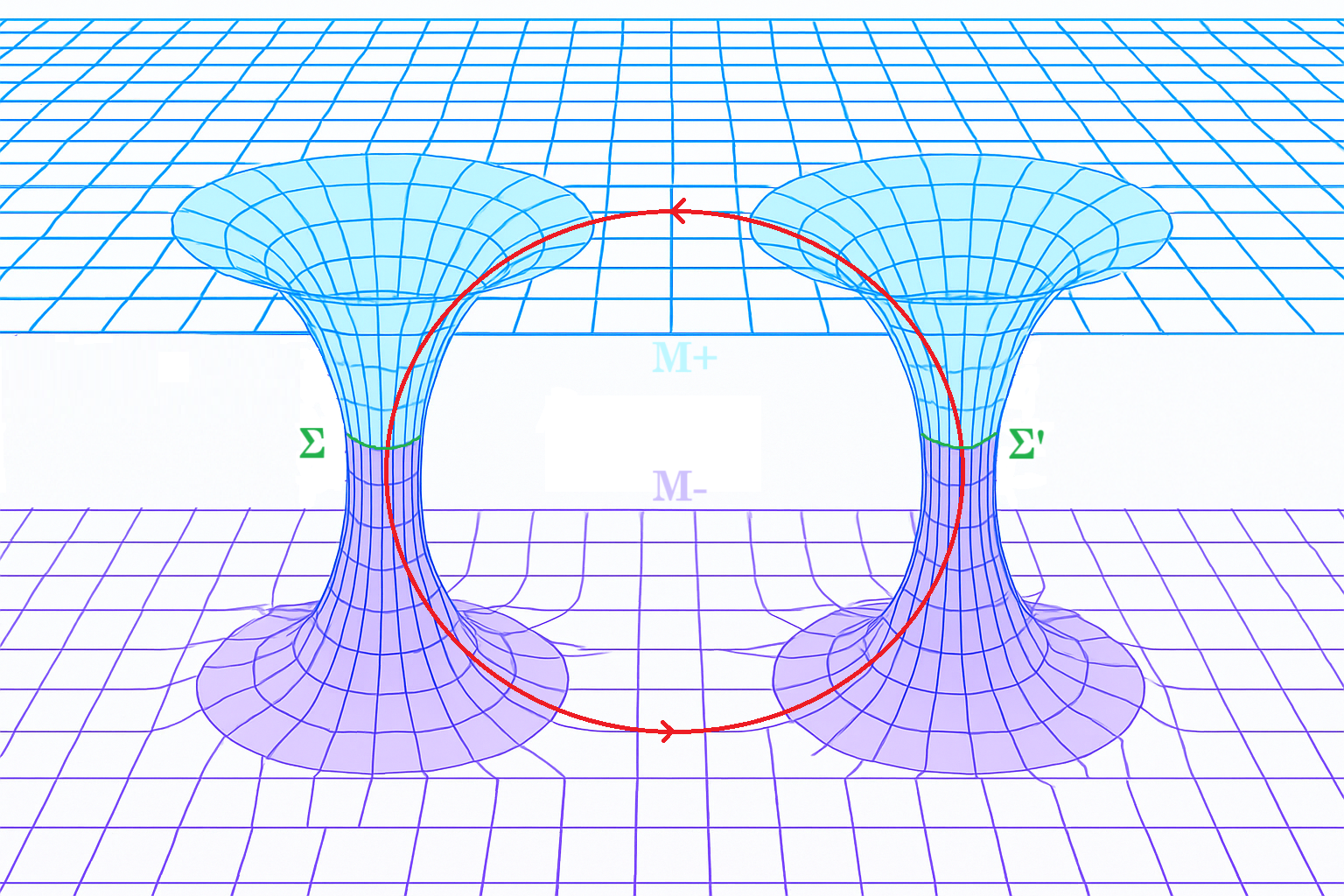}
\caption{Schematic representation of two $\mathcal{PT}$-symmetric bimetric wormholes. The regions $\mathcal{M}_+$ and $\mathcal{M}_-$ are identified at the throat via $\mathcal{PT}$ symmetry and connected by two distinct one-way bridges, extending the model of~\cite{koiran2024}. Each throat is centered on a junction hypersurface, denoted $\Sigma$ and $\Sigma'$, respectively. This geometry permits the existence of CTCs, shown in red, arising from the oriented causal flow and bimetric topology.}
\label{fig:CTC}
\end{figure}

Consider a spacetime in which the two regions $\mathcal{M}_+$ and $\mathcal{M}_-$ are identified at the throat via $\mathcal{PT}$ symmetry, forming a single sheet, as described in the model of~\cite{koiran2024}. In this model, a one-way bridge allows traversal from $\mathcal{M}_+$ to $\mathcal{M}_-$. To enable the formation of CTCs, we extend this model by introducing a second one-way bridge, such that the two regions are connected by two distinct bridges (see Figure~\ref{fig:CTC}). Each bridge is described by its respective metrics $g^{(+)}$ and $g^{(-)}$ in Eddington–Finkelstein coordinates\footnote{with units $c = 1$ and using the $(+,-,-,-)$ signature}:

\begin{itemize}
  \item \textbf{Incoming metric} \(g^{(+)}\) :
\begin{equation}\label{eq:metric_plus}
\mathrm{d}s^2_+ = \left(1 - \frac{\alpha}{r}\right) \mathrm{d}t^2 - \left(1 + \frac{\alpha}{r}\right) \mathrm{d}r^2 - \frac{2\alpha}{r} \, \mathrm{d}r\,\mathrm{d}t - r^2 (\mathrm{d}\theta^2 + \sin^2 \theta \, \mathrm{d}\phi^2),
\end{equation}

  \item \textbf{Outgoing metric} \(g^{(-)}\) :
\begin{equation}\label{eq:metric_minus}
\mathrm{d}s^2_- = \left(1 - \frac{\alpha}{r}\right) \mathrm{d}t^2 - \left(1 + \frac{\alpha}{r}\right) \mathrm{d}r^2 + \frac{2\alpha}{r} \, \mathrm{d}r\,\mathrm{d}t - r^2 (\mathrm{d}\theta^2 + \sin^2 \theta \, \mathrm{d}\phi^2).
\end{equation}
\end{itemize}

Each bridge enables one-way traversal between the regions $\mathcal{M}_+$ and $\mathcal{M}_-$ at the throat $r = \alpha$\footnote{Where $\alpha = 2m$ represents the Schwarzschild radius for a mass $m$.}, within this single sheet. These metrics, corresponding to the Schwarzschild solution in incoming and outgoing Eddington–Finkelstein coordinates, are non-degenerate at $r = \alpha$~\cite{koiran2024}. The metrics \eqref{eq:metric_plus} and \eqref{eq:metric_minus} describe traversals through the junction hypersurfaces $\Sigma$ and $\Sigma'$, potentially forming a closed timelike curve. This results in a bimetric construction where the global spacetime is a single sheet arising from the identification of the mirror regions $\mathcal{M}_+$ and $\mathcal{M}_-$ at the throat via $\mathcal{PT}$ symmetry, equipped with two distinct metrics for the incoming and outgoing regions. This approach resolves the gluing problem of the original Einstein–Rosen bridge~\cite{misner1973gravitation}, yielding a coherent spacetime with a bimetric structure.

\subsection{Effect of the $\mathrm{d}r,\mathrm{d}t$ sign on traversability in either direction}

These modified metrics ensure that each bridge is traversable in a single direction only, determined by the sign of the cross term $\mathrm{d}r\,\mathrm{d}t$. Specifically, under the metric $g^{(+)}$, the cross term $-\frac{2\alpha}{r}\,\mathrm{d}r\,\mathrm{d}t$ induces a causal flow orientation that favors geodesics directed toward the throat at $r = \alpha$, characterizing an \textit{incoming bridge}. Conversely, under $g^{(-)}$, the cross term $+\frac{2\alpha}{r}\,\mathrm{d}r\,\mathrm{d}t$ favors geodesics outgoing from the throat, defining an \textit{outgoing bridge}. In practice, this implies that an observer falling from $\mathcal{M}_+$ reaches the throat in finite coordinate time under $g^{(+)}$, but can only cross it in the incoming direction. Symmetrically, an observer in $\mathcal{M}_-$ can cross the throat of the second bridge toward $\mathcal{M}_+$, according to the metric $g^{(-)}$. Thus, while each wormhole bridge is one-way, the bimetric ensemble enables global bidirectional communication between the two spacetime regions.\\

Technically, the introduction of the cross term $\mathrm{d}r\,\mathrm{d}t$ eliminates the coordinate singularity at $r = \alpha$, corresponding to the Schwarzschild horizon. The throat no longer constitutes an infinite horizon but a regular passage: the modified metric remains \textit{non-degenerate at $r = \alpha$}. A traveler can thus reach and cross $r = \alpha$ in finite time according to the coordinate $t$, in contrast to the standard Schwarzschild black hole case, where such crossing would require infinite coordinate time~\cite{Oppenheimer1939}. In summary, each Eddington–Finkelstein-type metric ensures \textit{effective traversability} of the bridge in its permitted direction, made possible by the appropriate sign of the cross term $\mathrm{d}r\,\mathrm{d}t$.

\subsection{Event sequence of a traveler’s journey}
\label{sec:sec23}
We outline the sequence of events for a round-trip journey utilizing two one-way bridges, from the perspective of an observer using the global coordinate $t$, which is formally continuous across the junction but subject to an inversion due to $\mathcal{PT}$ symmetry:

\begin{enumerate}
  \item \textbf{Departure from the $\mathcal{M}_+$ region.} The traveler departs from the $\mathcal{M}_+$ region within the spacetime sheet at a coordinate time $t = t_{\text{dep}}$. They head toward the throat at $r = \alpha$ of the first wormhole. Under the metric $g^{(+)}$, their coordinate time $t$ evolves normally (future-oriented).

  \item \textbf{Crossing the first bridge ($\mathcal{M}_+ \to \mathcal{M}_-$).} The traveler reaches the junction surface $\Sigma$ (the throat) at $t = t_0$ and crosses it. Upon crossing, they enter the $\mathcal{M}_-$ region via the outgoing metric, within the same spacetime sheet identified through $\mathcal{PT}$ symmetry. \textit{At this instant, a time inversion occurs}, driven by $\mathcal{PT}$ symmetry and the bimetric structure: the crossing event, observed at $t = t_0$ in $\mathcal{M}_+$, initializes evolution in $\mathcal{M}_-$ such that the time coordinate is inverted relative to $\mathcal{M}_+$.

  Concretely, for a small proper time interval $\Delta \tau$ after crossing, the traveler is at $t = t_0 - \Delta t$ in $\mathcal{M}_-$, whereas they were at $t = t_0 + \Delta t$ in $\mathcal{M}_+$ just before (inversion $t \to -t$). In other words, they \textit{emerge in the past} relative to the initial reference frame in $\mathcal{M}_+$. If $t_{\text{sort}}$ denotes the arrival time in $\mathcal{M}_-$, then $t_{\text{sort}} < t_0$. This difference stems from the $\mathcal{T}$ symmetry at the throat crossing, combined with $\mathcal{P}$\footnote{The spatial symmetry $\mathcal{P}$ adjusts the spatial coordinates, e.g., $\vec{x} \to -\vec{x}$.}.

  \item \textbf{Transit through the $\mathcal{M}_-$ region.} The traveler continues their journey in the $\mathcal{M}_-$ region of the same spacetime sheet. Under the metric $g^{(-)}$, their coordinate time $t$ \textit{appears to decrease} with proper time (from the perspective of the initial $\mathcal{M}_+$ region). Nevertheless, local causality remains intact: their proper time continues to evolve normally (future-oriented), although they move toward increasingly earlier coordinate times relative to $\mathcal{M}_+$.

  \item \textbf{Return via the second bridge ($\mathcal{M}_- \to \mathcal{M}_+$).} Suppose that at a certain instant $t'_{\text{dep}}$ in $\mathcal{M}_-$, the traveler encounters the entrance of the second wormhole at $\Sigma'$. They enter it to return to the $\mathcal{M}_+$ region. This second crossing, governed by the metric $g^{(-)}$ on the $\mathcal{M}_-$ side and $g^{(+)}$ on the $\mathcal{M}_+$ side (the inverse configuration of the first bridge), induces \textit{a second time inversion} due to $\mathcal{PT}$ symmetry. The traveler emerges in $\mathcal{M}_+$ at an instant $t'_{\text{ret}}$. Owing to this second application of $\mathcal{PT}$ symmetry, it is possible for $t'_{\text{ret}} \leq t_{\text{dep}}$. In particular, if $t'_{\text{ret}} = t_{\text{dep}}$, the traveler returns precisely to the instant and location of their initial departure; if $t'_{\text{ret}} < t_{\text{dep}}$, they arrive \textit{before} their departure, completing a closed timelike curve.
\end{enumerate}

The $\mathcal{PT}$ symmetry, defined by the transformation $t \to -t$ and $\vec{x} \to -\vec{x}$, relates the metrics $g^{(+)}$ and $g^{(-)}$ at the throat. A particle crossing the throat from $\mathcal{M}_+$ to $\mathcal{M}_-$ appears to evolve backward in time: its position at $t = t_0 + \Delta t$ in $\mathcal{M}_+$ before crossing corresponds to $t = t_0 - \Delta t$ in $\mathcal{M}_-$ after~(\cite{koiran2024}). This apparent time inversion enables the formation of closed timelike curves by coupling two such bridges.

By identifying the sheets at $r = \alpha$, as proposed by Einstein and Rosen~\cite{einstein1935er}, we obtain a global bimetric structure equipped with two distinct metrics, $g^{(+)}$ and $g^{(-)}$. Unlike the original Einstein–Rosen bridge, where identification reduced to a single sheet, this structure yields a unique geometry. The $\mathcal{PT}$ symmetry carries physical significance: a particle crossing the throat could be interpreted as an antiparticle in the other sheet, in connection with $\mathcal{CPT}$ symmetry~\cite{koiran2024}. The crossing at the throat is facilitated by the continuity of the metric, ensured by Eddington-Finkelstein coordinates; however, a thin shell of exotic matter is present at the throat to satisfy the Einstein field equations, consistent with~\cite{guendelman2010,guendelman2016einstein,koiran2024}. 

\subsection{Construction of closed timelike geodesics without causality violation}

Consider a massive particle following a timelike geodesic in a bimetric spacetime where the regions $\mathcal{M}_+$ and $\mathcal{M}_-$ are identified at the throat via $\mathcal{PT}$ symmetry, forming a single sheet equipped with the metrics $g^{(+)}$ and $g^{(-)}$~\cite{koiran2024}. In our extension, these regions are connected by two one-way wormholes at the hypersurfaces $\Sigma$ and $\Sigma'$ ($r = \alpha$), enabling closed geodesics (see Figure~\ref{fig:CTC}). Crossing the throat $\Sigma$ at $t = t_0$ induces an apparent time inversion due to $\mathcal{PT}$ symmetry, such that the particle appears to move into the past in $\mathcal{M}_-$, while respecting local causality, as its proper time remains monotonic.

The metric $g^{(-)}$ is obtained by applying a $\mathcal{PT}$ transformation to $g^{(+)}$, i.e.,
\begin{equation}\label{eq:eq3}
g^{(-)}(x^\mu) = \mathcal{PT}[g^{(+)}(x^\mu)] = g^{(+)}(-t, -\vec{x}),
\end{equation}
inverting the time and spatial coordinates in $\mathcal{M}_-$ relative to $\mathcal{M}_+$. This transformation changes the sign of the cross term $\mathrm{d}r\,\mathrm{d}t$, reproducing the metrics \eqref{eq:metric_plus} and \eqref{eq:metric_minus}.\\

At the throat ($r=\alpha$), defined as the hypersurface $\Sigma$, the modified Eddington-Finkelstein coordinates \cite{koiran2021, koiran2024} ensure a continuous metric ($C^0$) between $\mathcal{M}_{+}$ and $\mathcal{M}_{-}$, supported by the $\mathcal{P}\mathcal{T}$-symmetric structure. A lightlike membrane of exotic matter is required at the throat to satisfy the Einstein field equations, introducing a surface stress-energy tensor $S_{\mu \nu}$, as proposed in \cite{guendelman2010, guendelman2016einstein, koiran2024}. The Lorentzian signature is preserved across $\Sigma$, avoiding any causal discontinuity. These properties enable crossing without metric discontinuity, except for the apparent time inversion of \(t\) due to \(\mathcal{PT}\) symmetry.\\

The particle follows a geodesic parameterized by its proper time $\tau$ along a trajectory $\gamma(\tau)$, satisfying the relativistic condition:
\begin{equation}\label{eq:eq4}
\exists \, \tau_1 < \tau_2 : \gamma(\tau_1) = \gamma(\tau_2)
\quad \text{et} \quad
g_{\mu\nu} \frac{\mathrm{d}\gamma^\mu}{\mathrm{d}\tau} \frac{\mathrm{d}\gamma^\nu}{\mathrm{d}\tau} < 0 \quad \forall \, \tau \in [\tau_1, \tau_2].
\end{equation}

The time coordinate $t$ is continuous across the crossing, but its derivative with respect to proper time changes sign:
\begin{itemize}
  \item Before the throat (under $g^{(+)}$): $\displaystyle \frac{\mathrm{d}t}{\mathrm{d}\tau} > 0$ (evolution toward the future).
  \item After the throat (under $g^{(-)}$): $\displaystyle \frac{\mathrm{d}t}{\mathrm{d}\tau} < 0$ (apparent evolution toward the past).
\end{itemize}

At the crossing ($\tau = \tau_0$), a time inversion occurs:
\begin{equation}\label{eq:eq5}
\left.\frac{\mathrm{d}t}{\mathrm{d}\tau}\right|_{\tau_0} = 0\,,
\end{equation}
marking the instant of time inflection, where the velocity vector is orthogonal to the vector $\partial_t$. The proper time $\tau$ evolves normally, but $t(\tau)$ decreases after the throat, with $\frac{\mathrm{d}t}{\mathrm{d}\tau} < 0$, while maintaining $\frac{\mathrm{d}\tau}{\mathrm{d}\tau} > 0$, preserving local causality\footnote{No local retrocausality or tachyons; only the coordinate $t$ undergoes a global inversion.}.\\

Note that the crossing time at the throat is finite for an external observer. In Eddington–Finkelstein coordinates, radial null geodesics reach $r = \alpha$ without a coordinate singularity. For instance, under $g^{(+)}$, an advanced time coordinate $\zeta = t + f(r)$\footnote{With $f(r) = \frac{\alpha}{c} \ln\left| \frac{r}{\alpha} - 1 \right|$.} remains finite at the throat~\cite{koiran2021, koiran2024}. This ensures that the durations $t_0 - t_{\text{dep}}$ (outward journey) and $t'_{\text{ret}} - t'_{\text{dep}}$ (return journey) are finite.

Thus, no time divergence prevents the round trip. The times $t_{\text{dep}}$, $t_0$, $t_{\text{sort}}$, $t'_{\text{dep}}$, and $t'_{\text{ret}}$ are organized according to the time inversions described in Section~\ref{sec:sec23}, ensuring a consistent chronology. The condition $t'_{\text{ret}} \leq t_{\text{dep}}$ allows for a closed loop, corresponding to a return at or before the departure time.\\

It is important to emphasize that completing a \textit{closed timelike curve} in this bimetric model requires the presence of a second one-way wormhole at $\Sigma'$, which enables the return crossing from $\mathcal{M}_-$ to $\mathcal{M}_+$, such that:
\begin{equation}\label{eq:eq6}
t'_{\rm ret} \le t_{\rm dep}
\end{equation}

The complete trajectory of the traveler is then composed as follows:
\begin{equation}\label{eq:eq7}
\gamma = \gamma_1 \cup \gamma_2 \quad \text{where} \quad \gamma_1 : \mathcal{M}_+ \rightarrow \mathcal{M}_-, \quad \gamma_2 : \mathcal{M}_- \rightarrow \mathcal{M}_+
\end{equation}

with the closure condition:
\begin{equation}\label{eq:eq8}
\gamma(\tau_{\rm final}) = \gamma(\tau_{\rm initial}), \quad \text{and} \quad
g^{(\pm)}_{\mu\nu} \frac{\mathrm{d}\gamma^\mu}{\mathrm{d}\tau} \frac{\mathrm{d}\gamma^\nu}{\mathrm{d}\tau} < 0
\end{equation}

indicating that the entire worldline is timelike throughout its trajectory.

\section{Novikov’s self-consistency principle and causal coherence}

The hypothesis of the existence of closed timelike curves in certain relativistic geometries inevitably raises the question of time paradoxes and the preservation of causality. In particular, if an individual is capable of intervening in their own past, this could theoretically lead to logical contradictions, such as the famous grandfather paradox, in which a time traveler prevents their own birth by eliminating one of their ancestors.

To avoid such inconsistencies, Igor Novikov and his collaborators proposed in the 1980s a principle of causal coherence, now known as the \textit{Novikov principle} or \textit{self-consistency principle}~\cite{Friedman1990,Novikov1990}. This principle stipulates that: ``if an event exists that would give rise to a paradox or any alteration of the past, then the probability of that event is zero.'' In other words, the laws of physics would only permit globally coherent evolutions, free from causal contradictions~\cite{Visser1995}.

This principle relies on the analysis of the Cauchy problem in spacetimes containing CTCs, where it is generally impossible to freely specify initial conditions on a spacelike hypersurface. Thus, any dynamic solution in such a framework must be compatible with a closed cyclic causal structure, without breaking coherence.\\

Applied to the case of a traversable wormhole generating a temporal loop, the Novikov principle implies that any attempt to alter the past will necessarily fail. If a traveler reappears in the past before their own departure, they will be integrated coherently into the history leading to their departure. Any attempt to prevent their own crossing must, therefore, inevitably fail or even paradoxically contribute to its realization.

This situation was paradigmatically illustrated by the famous Polchinski billiard paradox~\cite{Echeverria1991}, in which a billiard ball passing through a wormhole attempts to strike its younger self to prevent it from entering. Classical analyses have shown that there is always a self-consistent solution to this problem: the collision in the past slightly deflects the ball’s trajectory, but just enough to allow it to enter the wormhole in a manner that precisely generates that collision. In other words, the events in the loop adjust self-consistently, forming a closed configuration where each cause produces an effect compatible with its origin.\\

In the case of our bimetric $\mathcal{PT}$-symmetric two-wormhole model, a self-consistent \textit{time travel} scenario is possible since this principle applies similarly: the traveler crosses each throat in finite time, undergoes a time inversion during the passage due to $\mathcal{PT}$ symmetry, and can complete a temporal loop by combining two crossings. Adherence to the Novikov principle precludes any paradox: if a traveler emerges in the past, they cannot disrupt the origin of their own journey. Any attempt to alter the initial conditions must instead conform to them. For instance, an attempt to dissuade their ``younger self'' might, through a misunderstanding or psychological reaction, reinforce their determination to depart. Such scenarios reveal that free will is constrained within the framework of CTCs: individuals may make local decisions, but only those that fit into a globally coherent history are physically realizable. The $\mathcal{PT}$ symmetry of our model facilitates this coherence by ensuring that time inversions during crossings integrate into globally consistent trajectories, strengthening the application of the Novikov principle.\\

This form of retro-causal determinism, though counterintuitive, has the advantage of resolving paradoxes without mathematically prohibiting the existence of temporal loops. However, it conflicts with another major hypothesis: the \textit{chronology protection conjecture} formulated by Hawking~\cite{hawking1992}. This conjecture posits that the fundamental laws of physics prevent the formation of CTCs in practice by rendering unstable the geometries that support them.

More precisely, Hawking suggests that a destructive quantum mechanism\footnote{For example, a divergent feedback of vacuum energy.} would intervene at the onset of CTCs. Indeed, a quantum field subjected to a geometry containing a closed time region can generate a divergent vacuum energy density near the Cauchy horizon, such that :
\begin{equation}\label{eq:eq9}
\langle T_{\mu\nu} \rangle \rightarrow \infty\,.
\end{equation}
This instability would destroy the wormhole or prevent its complete formation.
Indeed, Hawking demonstrated that, in attempting to create a ``time machine'' with a wormhole, the vacuum electromagnetic field would experience a looped Doppler effect: virtual photons can traverse the temporal loop multiple times, amplifying with each pass. This phenomenon of \emph{vacuum polarization} induces an energy accumulation. For example, if a photon gains a frequency factor $\Lambda > 1$ (and thus energy $E$) per loop, its effective energy after $n$ loops becomes $E_n = \Lambda^n E_0$. The sum of these contributions diverges as $n \to \infty$:
\begin{equation}\label{eq:eq10}
\sum_{n=0}^{\infty} E_0\,\Lambda^n \to +\infty\,,
\end{equation}
for $\Lambda > 1$. In terms of the effective vacuum energy-momentum tensor $\langle T_{\mu\nu} \rangle$, this positive feedback translates into a \emph{divergence} of the energy density as soon as the first temporal loop becomes accessible. More precisely, Hawking’s semiclassical calculation indicates that an observer approaching the chronological horizon would measure a divergent energy density\footnote{$\rho \to \infty$, where \(\rho = \langle T_{00} \rangle\)} represents the local energy density) carried by vacuum fluctuations. Such an infinite energy barrier would cause the collapse or disintegration of the wormhole before a usable temporal loop could form. Ultimately, this conjecture\footnote{Often summarized by Hawking’s aphorism: ``Physics prevents time travel to preserve History.''} asserts that CTCs are simply excluded from realizable configurations in our universe.\\

However, recent studies suggest that mechanisms such as quantum gravity or specific symmetries, such as $\mathcal{PT}$ symmetry, could mitigate these instabilities by regularizing quantum fluctuations~\cite{maldacena2013}. These hypotheses require further analysis to confirm the viability of CTCs.\\

To date, the question of whether nature permits such configurations remains open. The Novikov principle provides an elegant conceptual solution to render CTCs physically admissible without paradoxes, while Hawking’s conjecture introduces a dynamic barrier to their realization.\\

In the present framework, we assume that the $\mathcal{PT}$-symmetric wormhole is sufficiently stable to allow the formation of a temporal loop. In this case, the Novikov self-consistency principle applies: each event in the temporal loop is both a consequence and a condition of another event in the cycle. For example, the traveler’s arrival in the past could trigger the events that later lead them to cross the wormhole, thus realizing a causally closed and mathematically coherent loop.\\

Ultimately, two possibilities emerge: either nature prohibits these configurations, in which case the model must be revised; or it permits them, but necessarily in a self-consistent manner. In this paper, we adopt the latter hypothesis, considering that the CTCs enabled by the bimetric structure of our model are compatible with the Novikov principle and thus free from any causal contradiction.\\

Next, we will examine the dynamic viability of this configuration in light of the Einstein field equations and energy conditions.

\section{Einstein field equations and energy conditions}

\subsection{Local validity of the Einstein equations}

The metrics \eqref{eq:metric_plus} and \eqref{eq:metric_minus}, introduced earlier for the incoming and outgoing sheets of the modified bridge, are regular reformulations of the Schwarzschild metric for a mass \(m = \alpha / 2\) in Eddington–Finkelstein coordinates. For \(r > \alpha\), they locally satisfy the Einstein equations in vacuum\footnote{\(R_{\mu\nu} = 0 \implies G_{\mu\nu} = 0\)} without a cosmological constant. Physically, each sheet represents the gravitational exterior of a point-like object of mass \(m\), with the spacetime curved by the central mass but devoid of distributed material content.

\subsection{A geometrically regular throat}
\label{sec:Abs_Source_Locale}

The junction of the two sheets at $r=\alpha$ is facilitated by the geometric continuity of the metric, enabled by Eddington-Finkelstein coordinates \cite{koiran2024}. A lightlike membrane of exotic matter, in the form of a surface stress-energy tensor $S_{\mu \nu}$, is required at this junction to ensure the physical consistency of the spacetime, consistent with models proposed in \cite{guendelman2010, guendelman2016einstein, Bronnikov2018, koiran2024}. This exotic matter violates the null energy condition (NEC), a standard requirement for traversable wormholes to maintain an open throat \cite{morris1988,Visser1995}. The bulk regions (\(\eta = 0\) in~\cite{koiran2024}) have zero stress-energy, but the non-zero \( S_{\mu\nu} \) at \( r = \alpha \) may violate classical energy conditions like the NEC and weak energy condition (WEC), raising questions about dynamic stability that warrant further study~\cite{Visser1995}.\\

The primary objective of this paper is to demonstrate the theoretical feasibility of closed timelike curves (CTCs) through the coupling of two \(\mathcal{PT}\)-symmetric wormholes, leveraging the bimetric structure and Eddington-Finkelstein coordinates to ensure traversability and causal consistency~\cite{koiran2024}. While the presence of a lightlike membrane of exotic matter at the throat (\( r = \alpha \)) is necessary to satisfy the Einstein field equations and account for the discontinuity in the metric derivatives via Israel’s junction conditions~\cite{israel1966}, detailed calculations of the surface stress-energy tensor and the extrinsic curvature jump are beyond the scope of this study. Such computations, which are standard in wormhole models~\cite{ guendelman2010,guendelman2016einstein,morris1988,Visser1995}, will be addressed in a future article, allowing the present work to focus on the novel geometric and causal implications of the coupled wormhole system.

\subsection{Possible extensions of the model}

Building on the current model, which incorporates a lightlike membrane of exotic matter at the throat (as discussed in Section~\ref{sec:Abs_Source_Locale}), future extensions could explore more complex topologies to enhance its theoretical scope. One promising direction involves networks of interconnected wormholes, potentially linked to large-scale quantum entanglement effects, as proposed in the context of entangled black holes~\cite{maldacena2013}. Such configurations might offer insights into quantum gravity or the role of \(\mathcal{PT}\) symmetry in broader cosmological frameworks. These extensions would require detailed numerical analyses to evaluate their physical feasibility and stability, building on the foundational considerations already addressed~\cite{Visser1995}.

\section{Quantization and \(\mathcal{PT}\) symmetry}

\subsection{Effective theory of a scalar field crossing the \(\mathcal{PT}\)-symmetric wormhole}
\label{sec:theorie_effective}

We develop here an effective theory to describe a complex scalar field crossing the bimetric \(\mathcal{PT}\)-symmetric wormhole, drawing inspiration from studies of scalar fields in wormholes~\cite{Bronnikov2018} and the properties of \(\mathcal{PT}\)-symmetric systems in quantum mechanics~\cite{bender1998, bender2002}. This approach relies on constructing a \(\mathcal{PT}\)-invariant action, from which we derive equations of motion consistent with the geometry of the regions \(\mathcal{M}_+\) and \(\mathcal{M}_-\).

\subsubsection{Geometry and \(\mathcal{PT}\) symmetry}

We have seen that our wormhole model is described by two metrics, corresponding to modified Eddington–Finkelstein coordinates, ensuring one-way traversability and geometric regularity at the throat at \(r = \alpha\). The metrics associated with the regions \(\mathcal{M}_+\) (incoming) and \(\mathcal{M}_-\) (outgoing) are given by \eqref{eq:metric_plus} and \eqref{eq:metric_minus}, where the sign change of the cross term \(\mathrm{d}r \mathrm{d}t\) reflects the \(\mathcal{PT}\) symmetry, defined by the transformation \(t \to -t\), \(\vec{x} \to -\vec{x}\). This transformation links the two sheets, ensuring that for a point \(x = (t, r, \theta, \phi)\) in \(\mathcal{M}_+\), the corresponding point in \(\mathcal{M}_-\) is \(\mathcal{PT}x = (-t, r, \pi - \theta, \phi + \pi)\), adjusted to preserve the spherical topology.\\

As considered in Section~\ref{sec:Abs_Source_Locale}, the throat at \( r = \alpha \) hosts a thin shell of exotic matter that affects the spacetime geometry, ensuring a continuous metric via Eddington-Finkelstein coordinates. The metric continuity facilitates a smooth geometric transition between \(\mathcal{M}_+\) and \(\mathcal{M}_-\).\\

In the presence of a lightlike membrane of exotic matter at the throat, as suggested by \cite{guendelman2010,guendelman2016einstein,koiran2024}, the surface stress-energy tensor $S_{\mu \nu}$ affects the geometry, ensuring a continuous metric via Eddington-Finkelstein coordinates. However, we assume minimal coupling between this shell and the complex scalar field \(\phi\), such that no direct interaction occurs at the throat\footnote{In general relativity, a minimally coupled scalar field interacts with the gravitational field only through the spacetime metric, with no direct coupling to the curvature~\cite{BirrellDavies1982, Wald1984}. This is reflected in the Klein-Gordon equation in curved spacetime, obtained via the minimal substitution \( \eta_{ab} \rightarrow g_{ab} \), leading to \(\nabla^a \nabla_a \phi - m^2 \phi = 0\) (§4.3, eq.~4.3.9 of \cite{Wald1984}). Any additional coupling term involving the Ricci scalar, such as \( \alpha R \phi \), constitutes a non-minimal modification. As shown in Appendix D of \cite{Wald1984} (eqs.~D.11–D.13), such terms are introduced specifically to ensure conformal invariance of the scalar field equation.}. This assumption is consistent with standard treatments of scalar fields in curved spacetimes, where fields couple to gravity solely through the metric, without additional interactions with localized matter sources~\cite{BirrellDavies1982}. The thin shell’s purely gravitational role, as in Visser’s wormhole models~\cite{Visser1995}, ensures that the scalar field dynamics remain governed by the bulk action (\ref{eq:total_action}--\ref{eq:action_minus}) in \(\mathcal{M}_+\) and \(\mathcal{M}_-\). The \(\mathcal{PT}\)-symmetric junction condition (\ref{eq:junction_condition_exact}) thus remains valid, preserving the \(\mathcal{PT}\) invariance and pseudo-unitary structure of the quantum dynamics, as detailed in next sections.

\subsubsection{Construction of the action}
We define two complex scalar fields, \(\phi_+\) on \(\mathcal{M}_+\) with the metric \(g^{(+)}\), and \(\phi_-\) on \(\mathcal{M}_-\) with the metric \(g^{(-)}\), related by \(\mathcal{PT}\) symmetry. The \(\mathcal{PT}\) transformation acts on a complex scalar field as follows: \(\phi(t, \vec{x}) \to \phi^*(-t, -\vec{x})\)\footnote{where \(*\) denotes complex conjugation}. Thus, for a point \(x\) in \(\mathcal{M}_-\), we impose:

\begin{equation}
\phi_-(x) = \phi_+^*(\mathcal{PT}^{-1}x), \label{eq:PT_constraint}
\end{equation}

where \(\mathcal{PT}^{-1}x\) is the point in \(\mathcal{M}_+\) corresponding to \(x\) via the \(\mathcal{PT}\) inversion. This relation ensures that \(\phi_-\) is not an independent degree of freedom but is determined by \(\phi_+\), guaranteeing \(\mathcal{PT}\) invariance.\footnote{The use of complex scalar fields is motivated by the nature of the \(\mathcal{PT}\) transformation, which includes complex conjugation for fields. Under \(\mathcal{PT}\) transformation, a complex scalar field \(\phi\) transforms to \(\phi^*(-t, -\vec{x})\), preserving the structure of the action and the junction condition. A real field, however, would need to satisfy \(\phi(-t, -\vec{x}) = \phi(t, \vec{x})\), imposing stricter symmetry constraints and complicating the dynamics at the throat. Complex fields thus offer greater flexibility for modeling the wormhole crossing while respecting \(\mathcal{PT}\) symmetry.}\\

The total action for the complex scalar fields \(\phi_+\) and \(\phi_-\) in the bimetric \(\mathcal{PT}\)-symmetric wormhole is defined as the sum of contributions from the regions \(\mathcal{M}_+\) and \(\mathcal{M}_-\)~\cite{bender1998, Wald1984}:
\begin{equation}
S = S_+ + S_-, \label{eq:total_action}
\end{equation}

where:

\begin{equation}
S_+ = \int_{\mathcal{M}_+} \mathrm{d}^4x \sqrt{-g^{(+)}} \left( g_+^{\mu\nu} \partial_\mu \phi_+ \partial_\nu \phi_+^* - m^2 \phi_+ \phi_+^* \right),
\label{eq:action_plus}
\end{equation}

and similarly for \(\phi_-\) on \(\mathcal{M}_-\):

\begin{equation}
S_- = \int_{\mathcal{M}_-} \mathrm{d}^4x \sqrt{-g^{(-)}} \left( g_-^{\mu\nu} \partial_\mu \phi_- \partial_\nu \phi_-^* - m^2 \phi_- \phi_-^* \right), \label{eq:action_minus}
\end{equation}

with \(m\) the mass of the scalar field, and \(\sqrt{-g^{(\pm)}}\) the determinant of the metric on each sheet. This action includes a standard kinetic term and a simple mass potential, which represents the potential energy associated with the field’s mass \(m\), avoiding complications from ghost fields sometimes required in wormholes~\cite{Bronnikov2018}.\\

At the throat \(r = \alpha\), a junction condition is necessary to ensure the continuity of the field across the regions \(\mathcal{M}_+\) and \(\mathcal{M}_-\). We impose the exact condition derived from the \(\mathcal{PT}\) transformation:
\begin{equation}
\phi_-(t, r, \theta, \phi)\big|_{r=\alpha} = \phi_+^*(-t, r, \pi-\theta, \phi+\pi)\big|_{r=\alpha}. \label{eq:junction_condition_exact}
\end{equation}

This condition fully incorporates the \(\mathcal{PT}\) transformation, including the temporal inversion (\(t \to -t\)) and spatial inversions (\(\theta \to \pi - \theta\), \(\phi \to \phi + \pi\)), along with the complex conjugation. It ensures rigorous compliance with \(\mathcal{PT}\) symmetry at the throat. While this introduces additional complexity due to the angular and temporal transformations, it provides a more precise description of the field dynamics across the wormhole throat.

\subsubsection{Solutions of the scalar field under \(\mathcal{PT}\) symmetry}

To rigorously describe the behavior of a complex scalar field traversing the \(\mathcal{PT}\)-symmetric bimetric wormhole, we develop here the solutions to the Klein-Gordon equations in the regions \(\mathcal{M}_+\) and \(\mathcal{M}_-\), adhering to the exact junction condition at the throat \(r = \alpha\). This development employs the method of separation of variables to formulate the solutions and derive the radial equations, followed by a detailed analysis of the implications of \(\mathcal{PT}\) symmetry and a semi-classical quantization approach.

\subsubsection*{Equations of Motion}

Consider a complex scalar field \(\phi\), with \(\phi_+\) defined in \(\mathcal{M}_+\) (metric \(g^{(+)}\)) and \(\phi_-\) in \(\mathcal{M}_-\) (metric \(g^{(-)}\)), connected at the throat \(r = \alpha\). The \(\mathcal{PT}\) symmetry imposes the condition \ref{eq:junction_condition_exact}.\\

For simplicity, we assume spherical symmetry, eliminating angular dependence:
\begin{equation}
\phi_-(t, r)\big|_{r=\alpha} = \phi_+^*(-t, r)\big|_{r=\alpha}.\label{eq:junction_condition_exact_sym}
\end{equation}

The equations of motion are obtained by varying the action \eqref{eq:total_action} with respect to \(\phi_+^*\) and \(\phi_-^*\). On each sheet, this leads to the Klein-Gordon equation in the corresponding curved spacetime:
\begin{itemize}
    \item On \( \mathcal{M}_+ \) :
    \begin{equation}
    \Box_+ \phi_+ + m^2 \phi_+ = 0, \label{eq:KG_plus}
    \end{equation}
    where \( \Box_+ = g_+^{\mu\nu} \nabla_\mu \nabla_\nu \) is the d’Alembertian operator associated with \( g^{(+)} \). Neglecting the Christoffel connection terms for simplification\footnote{Valid for a scalar field without non-minimal coupling~\cite{BirrellDavies1982}.}, this becomes:
    \begin{equation}
    g_+^{\mu\nu} \partial_\mu \partial_\nu \phi_+ + m^2 \phi_+ = 0. \label{eq:KG_plus_simplified}
    \end{equation}
    \item On \( \mathcal{M}_- \) :
    \begin{equation}
    \Box_- \phi_- + m^2 \phi_- = 0, \label{eq:KG_minus}
    \end{equation}
    or similarly:
    \begin{equation}
    g_-^{\mu\nu} \partial_\mu \partial_\nu \phi_- + m^2 \phi_- = 0. \label{eq:KG_minus_simplified}
    \end{equation}
\end{itemize}

Detailed calculations are provided in Appendices~\ref{app:kg_derivation} and \ref{app:christoffel}.

\subsubsection*{Formulation of the scalar field solutions under \(\mathcal{PT}\) symmetry}

We can solve these equations by expressing the solutions as superpositions of oscillatory modes:

\begin{equation}
\phi_\pm(t, r) = \int_{-\infty}^{\infty} \mathrm{d}\omega \, a_\pm(\omega) e^{-i \omega t} R_\omega^\pm(r),
\end{equation}

where \(a_\pm(\omega)\) are the amplitudes of the modes with frequency \(\omega\), and \(R_\omega^\pm(r)\) are the radial functions associated with \(\mathcal{M}_+\) and \(\mathcal{M}_-\), solutions to the radial equations derived from the Klein-Gordon equations.\footnote{The equations with the index \(\pm\) correspond to the regions \(\mathcal{M}_+\) and \(\mathcal{M}_-\).}\\

Let us now apply the junction condition at the throat, given by \ref{eq:junction_condition_exact_sym}, at \(r = \alpha\). The left-hand side of the equation yields:
\begin{equation}
    \phi_-(t, \alpha) = \int_{-\infty}^{\infty} \mathrm{d}\omega \, a_-(\omega) e^{-i \omega t} R_\omega^-(\alpha).
\end{equation}

The right-hand side leads to:
\begin{equation}
    \phi_+^*(-t, \alpha) = \int_{-\infty}^{\infty} \mathrm{d}\omega \, a_+^*(-\omega) e^{-i \omega t} \left( R_{-\omega}^+(\alpha) \right)^*.
\end{equation}

Thus, \ref{eq:junction_condition_exact_sym} allows us to obtain:
\begin{equation}
    \int_{-\infty}^{\infty} \mathrm{d}\omega \, a_-(\omega) e^{-i \omega t} R_\omega^-(\alpha) = \int_{-\infty}^{\infty} \mathrm{d}\omega \, a_+^*(-\omega) e^{-i \omega t} \left( R_{-\omega}^+(\alpha) \right)^*.
\end{equation}

This leads, for each \(\omega\), to:
\begin{equation}\label{eq:eq_radiale_1}
    a_-(\omega) R_\omega^-(\alpha) = a_+^*(-\omega) \left( R_{-\omega}^+(\alpha) \right)^*.
\end{equation}

Assuming that the radial functions at the throat are real and identical, we obtain:
\begin{equation}
    R_\omega^-(\alpha) = R_\omega^+(\alpha) = R_\omega(\alpha).
\end{equation}  

Furthermore, under the assumption of a slow temporal evolution at the throat (\(\omega \ll 1\))\footnote{That is, for fields varying slowly in time.}, the linear terms in \(\omega\) in the radial equation (detailed in Appendix~\ref{app:radial_eq}) become negligible compared to those in \(\omega^2\). The radial equation is invariant under \(\omega \to -\omega\), allowing us to deduce that \(R_\omega(\alpha) \approx R_{-\omega}(\alpha)\), which is relevant for configurations with low temporal dynamics.\\

The relation \ref{eq:eq_radiale_1} then leads to the condition relating the mode amplitudes in \(\mathcal{M}_-\) to those in \(\mathcal{M}_+\):
\begin{equation}\label{eq:condition}
    a_-(\omega) = a_+^*(-\omega).
\end{equation}

\subsubsection*{Semi-classical quantization under \(\mathcal{PT}\) symmetry}

In the framework of semi-classical quantization, the classical amplitudes \(a_+(\omega)\) and \(a_-(\omega)\), along with their complex conjugates \(a_+^*(\omega)\) and \(a_-^*(\omega)\), become operators acting in Fock space. We define:

\begin{itemize}
    \item \(\hat{a}_+(\omega)\), the annihilation operator for modes of frequency \(\omega\) in \(\mathcal{M}_+\),
    \item \(\hat{a}_+^\dagger(\omega)\), the corresponding creation operator in \(\mathcal{M}_+\),
    \item \(\hat{a}_-(\omega)\), the annihilation operator in \(\mathcal{M}_-\),
    \item \(\hat{a}_-^\dagger(\omega)\), the creation operator in \(\mathcal{M}_-\).
\end{itemize}

These operators satisfy the canonical commutation relations for bosonic fields:

\begin{equation}
[\hat{a}_+(\omega), \hat{a}_+^\dagger(\omega')] = \delta(\omega - \omega'), \quad [\hat{a}_-(\omega), \hat{a}_-^\dagger(\omega')] = \delta(\omega - \omega'),
\end{equation}

with all other commutators vanishing.\footnote{For example, \([\hat{a}_+(\omega), \hat{a}_-(\omega')] = 0\).}\\

The condition \ref{eq:condition} becomes:
\begin{equation}\label{eq:condition_2}
    \hat{a}_-(\omega) = \hat{a}_+^\dagger(-\omega).
\end{equation}

This relation indicates that the annihilation of a particle in \(\mathcal{M}_-\) with frequency \(\omega\) corresponds to the creation of a particle in \(\mathcal{M}_+\) with frequency \(-\omega\)\footnote{This transformation is standard in quantum field theory, transitioning from a classical description (waves) to a quantum description (particles). The \(\mathcal{PT}\) condition imposes a specific structure, consistent with the symmetry of the system.}.\\

Then, we can deduce the following relation:
\begin{equation}\label{eq:condition_3}
    \hat{a}_+(-\omega) = \hat{a}_-^\dagger(\omega).
\end{equation}

 This physically reflects:
\begin{itemize}
    \item A symmetry between particles and antiparticles,
    \item A relationship between modes of opposite frequencies in a \(\mathcal{PT}\)-symmetric system, such as a wormhole with two symmetric sides.
\end{itemize}

The scalar fields then become quantum operators:
\begin{equation}
\hat{\phi}_+(t, r) = \int_{-\infty}^{\infty} \mathrm{d}\omega \left[ \hat{a}_+(\omega) e^{-i \omega t} R_\omega^+(r) + \hat{a}_+^\dagger(\omega) e^{i \omega t} \left( R_\omega^+(r) \right)^* \right],
\end{equation}

\begin{equation}
\hat{\phi}_-(t, r) = \int_{-\infty}^{\infty} \mathrm{d}\omega \left[ \hat{a}_-(\omega) e^{-i \omega t} R_\omega^-(r) + \hat{a}_-^\dagger(\omega) e^{i \omega t} \left( R_\omega^-(r) \right)^* \right].
\end{equation}

Using the relations \ref{eq:condition_2} and \ref{eq:condition_3}, this yields:
\begin{equation}
\hat{\phi}_-(t, r) = \int_{-\infty}^{\infty} \mathrm{d}\omega \left[ \hat{a}_+^\dagger(-\omega) e^{-i \omega t} R_\omega^-(r) + \hat{a}_+(-\omega) e^{i \omega t} \left( R_\omega^-(r) \right)^* \right].
\end{equation}

These quantum fields satisfy the Klein-Gordon equations in their respective regions, and the junction condition at \(r = \alpha\) is upheld at the operator level, ensuring the consistency of the quantum dynamics across the wormhole throat.

\subsubsection*{Implications for \(\mathcal{PT}\) symmetry}

The condition \eqref{eq:condition_2} reflects the \(\mathcal{PT}\) symmetry, where a particle described by the complex scalar field \(\phi_-\) in \(\mathcal{M}_-\) is annihilated and a corresponding particle described by \(\phi_+\) in \(\mathcal{M}_+\) is created. Although the Hamiltonian is non-Hermitian, it commutes with the \(\mathcal{PT}\) operator, ensuring a real energy spectrum through pseudo-unitarity. This is defined by a modified inner product \(\langle \psi | \mathcal{CPT} | \psi \rangle\), where \(\psi\) is a quantum state in the Hilbert space associated with \(\hat{\phi}_+\) and \(\hat{\phi}_-\), and \(\mathcal{CPT}\) incorporates the \(\mathcal{PT}\) operator along with a linear operator \(C\) ensuring positivity~\cite{bender1998,bender2002}.\\

This development provides a rigorous foundation for analyzing scalar fields in this wormhole, consistent with the bimetric geometry and the envisioned closed timelike curves.

\subsubsection*{Quantization and \(\mathcal{PT}\) invariance}

The structure of the Klein-Gordon equations \eqref{eq:KG_plus} and \eqref{eq:KG_minus}, combined with the \(\mathcal{PT}\) constraint \eqref{eq:PT_constraint} and the exact junction condition under spherical symmetry \eqref{eq:junction_condition_exact_sym}, ensures compatibility with semi-classical quantization. These equations are linear, and the approximate boundary condition at the throat \(r = \alpha\), given by \eqref{eq:junction_condition_exact_sym}, is well-defined, facilitating the construction of quantum states in curved spacetime~\cite{misner1973gravitation}. The \(\mathcal{PT}\) symmetry imposes a critical structure, as evidenced by the classical mode amplitude relation \eqref{eq:condition}, \(a_-(\omega) = a_+^*(-\omega)\), and its operator counterpart \eqref{eq:condition_2}, suggesting a pseudo-unitary framework akin to non-Hermitian \(\mathcal{PT}\)-symmetric systems~\cite{bender1998}.\\

In this \(\mathcal{PT}\)-symmetric context, states on \(\mathcal{M}_+\) and \(\mathcal{M}_-\), denoted \(|\psi_+\rangle\) and \(|\mathcal{PT} \psi_+\rangle\), form conjugate pairs. The inner product is redefined as \(\langle \psi_+ | \mathcal{CPT} | \psi_+ \rangle = \langle \mathcal{PT} \psi_+ | \psi_+ \rangle\), where the operator \(\mathcal{CPT}\) incorporates \(C\), a linear operator ensuring positivity~\cite{bender2002}. This modified inner product is related to the pseudo-unitary metric \(\eta\), a positive Hermitian operator introduced in the context of pseudo-unitarity, such that \(\eta\) can be expressed in terms of \(\mathcal{CPT}\) and additional symmetry operators~\cite{bender2002}. This structure resolves potential issues with negative-energy modes ("ghosts") that might arise from conjugation across sheets~\cite{Kuntz2024}. For instance, a mode in \(\mathcal{M}_-\) that appears to have negative energy from the \(\mathcal{M}_+\) perspective is the \(\mathcal{PT}\)-transformed counterpart of a positive-energy mode in \(\mathcal{M}_+\), consistent with \eqref{eq:condition_2}.\\

The evolution operator across the wormhole satisfies a pseudo-unitarity condition, \(U^\dagger \eta U = \eta\), where \(\eta\) is an adapted metric preserving a generalized norm. This property, validated in models like \(H = p^2 + i x^3\)~\cite{bender1998} and extended to quadratic gravity~\cite{Kuntz2024}, ensures a real energy spectrum and coherent quantization. The exact junction condition and mode relations reinforce this framework, providing a robust basis for quantum field dynamics in the wormhole.

\subsection{Discussion and perspectives}

This effective theory sidesteps the need for ghost fields common in classical wormhole models~\cite{Bronnikov2018}, leveraging the \(\mathcal{PT}\)-symmetric nature of complex scalar fields. The use of complex fields, where the relation \eqref{eq:PT_constraint} and the junction condition under spherical symmetry \eqref{eq:junction_condition_exact_sym}, is justified by the requirement of complex conjugation under \(\mathcal{PT}\) transformations. Real fields, while feasible, would impose stricter symmetry constraints (e.g., \(\phi(-t, -\vec{x}) = \phi(t, \vec{x})\)), complicating dynamics at the throat without offering the flexibility of complex conjugation.\\

Our approach aligns with \(\mathcal{PT}\)-symmetric wormhole models~\cite{koiran2024}, enabling traversability through a combination of \(\mathcal{PT}\) symmetry and a lightlike membrane of exotic matter at the throat, as explored in~\cite{Bronnikov2018}. The exact junction condition \eqref{eq:junction_condition_exact_sym} provides a precise description of field dynamics at the throat, capturing time inversion and complex conjugation effects. However, its reliance on spherical symmetry limits its generality; extending the analysis to non-spherical configurations remains a future challenge.\\

Quantum stability poses a significant question, particularly in light of Hawking’s chronology protection conjecture~\cite{hawking1992}, which predicts vacuum energy divergence near closed timelike curves. Here, \(\mathcal{PT}\) symmetry may regulate fluctuations via pseudo-unitarity, with conjugate modes potentially canceling divergent terms, a mechanism observed in \(\mathcal{PT}\)-symmetric theories~\cite{bender1998}. A semi-classical computation of the energy-momentum tensor \(\langle T_{\mu\nu} \rangle\) is essential to validate this hypothesis.\\

Future investigations could focus on explicit eigenmode solutions using the exact junction condition or explore quantum fluctuations at the throat. The framework, defined by equations \eqref{eq:total_action}, \eqref{eq:junction_condition_exact_sym}, \eqref{eq:KG_plus}, and \eqref{eq:KG_minus}, establishes a quantifiable model for a scalar field in a \(\mathcal{PT}\)-symmetric bimetric wormhole, inviting further study into its quantum and cosmological implications.

\section{Conclusion}
\label{sec:conclusion}

We have explored an innovative model of a modified Einstein-Rosen wormhole, characterized by a bimetric structure and \(\mathcal{PT}\) symmetry, enabling one-way traversability and the formation of closed timelike curves. Through the use of modified Eddington-Finkelstein coordinates~\cite{koiran2021, koiran2024}, the throat at \( r = \alpha \) is made geometrically regular, resolving the topological gluing issues of the original Einstein-Rosen bridges~\cite{einstein1935er} by ensuring traversability in finite time, albeit with the inclusion of a lightlike membrane of exotic matter at the throat to maintain physical consistency. The \(\mathcal{PT}\) symmetry, defined by the inversions \( t \to -t \) and \(\vec{x} \to -\vec{x}\), imparts a unique geometric richness by identifying the regions \(\mathcal{M}_+\) and \(\mathcal{M}_-\) into a single spacetime sheet~\cite{koiran2024}. In our extension, the introduction of two \(\mathcal{PT}\)-symmetric wormholes allows for one-way crossings within this sheet, forming a closed temporal loop through successive time inversions between \(\mathcal{M}_+\) and \(\mathcal{M}_-\).\\

The analysis of geodesics has shown that CTCs are feasible without violating local causality, owing to metric continuity and the apparent time inversion during throat crossing. The Novikov self-consistency principle ensures the absence of paradoxes by requiring that any intervention in the past integrates coherently into the global history. For instance, a traveler returning before their departure cannot alter their own journey, as illustrated by scenarios such as the Polchinski billiard paradox \cite{Echeverria1991}. However, Hawking’s chronology protection conjecture poses a challenge: the potential divergence of vacuum energy near the Cauchy horizon could destabilize the wormhole \cite{hawking1992}. The effective theory developed in this paper, by incorporating \(\mathcal{PT}\) symmetry and a pseudo-unitary framework, offers a solution to stabilize quantum fluctuations. This approach ensures a real energy spectrum and resolves issues with negative-energy modes, building on the foundational work of Bender and Kuntz~\cite{bender1998,Kuntz2024}. Specifically, the \(\mathcal{PT}\)-symmetric junction condition and the quantization scheme developed in Section \ref{sec:theorie_effective} offer a mechanism to control vacuum fluctuations, potentially enabling the formation of closed timelike curves. This hypothesis warrants further analysis.\\

Indeed, the effective scalar field theory, detailed in Section~\ref{sec:theorie_effective} and Appendix~\ref{app:radial_eq}, provides a framework for analyzing a complex scalar field in this bimetric geometry. By defining an action that respects \(\mathcal{PT}\) invariance across both regions, we derived Klein-Gordon equations coupled through a precise junction condition at the throat, \(\phi_-(t, r) = \phi_+^*(-t, r)\), under spherical symmetry. The calculation of the D'Alembertian operator \(\Box_+\) for the scalar field \(\phi(t, r) = e^{-i \omega t} R_\omega(r)\) in \(\mathcal{M}_+\) yields the radial Klein-Gordon equation \eqref{eq_radial_positive_region} of the Appendix~\ref{app:radial_eq}. The analogous equation for \(\mathcal{M}_-\) confirms the consistency of the \(\mathcal{PT}\)-symmetric framework, with mode amplitudes related by \(a_-(\omega) = a_+^*(-\omega)\). This relation implies that a particle created in \(\mathcal{M}_-\) with frequency \(\omega\) corresponds to an antiparticle annihilated in \(\mathcal{M}_+\) with frequency \(-\omega\), or conversely, a particle created in \(\mathcal{M}_+\) with \(\omega\) appears as an antiparticle in \(\mathcal{M}_-\) with \(-\omega\). This reflects the \(\mathcal{PT}\) symmetry at the quantum level, where complex conjugation and temporal inversion link modes of opposite frequencies across the throat crossing. Physically, this indicates that the passage between regions transforms the quantum states of the scalar field, preserving the system's pseudo-unitarity and ensuring a real energy spectrum, consistent with non-Hermitian \(\mathcal{PT}\)-symmetric systems~\cite{bender1998,Kuntz2024}. This transformation aligns with the apparent causal inversion at the throat, where a particle crossing from \(\mathcal{M}_+\) to \(\mathcal{M}_-\) appears to evolve backward in time in the \(\mathcal{M}_+\) reference frame, supporting the feasibility of CTCs in the bimetric configuration. Semiclassical quantization, supported by this pseudo-unitarity, mitigates the risk of negative-energy modes, reinforcing the quantum coherence of the model.\\

Theoretical validation could involve detecting signatures in gravitational wave signals or cosmic microwave background fluctuations, though such observations depend on future technological advancements.\\

Quantum stability against Hawking’s conjecture requires explicit calculations, such as evaluating the energy-momentum tensor \(\langle T_{\mu \nu} \rangle\) in the bimetric geometry using semiclassical methods. Additionally, to further explore the scalar field’s behavior near the critical boundary \(r = \alpha\), numerical solutions to the radial Klein-Gordon equations derived in Appendix~\ref{app:radial_eq}, employing techniques such as separation of variables or finite difference methods, could be important for identifying stable eigenmodes or novel physical phenomena in the bimetric geometry. Furthermore, extensions involving throats supported by thin shells or complex topologies, such as wormhole networks, could enrich the model~\cite{maldacena2013}. These research directions, combined with semiclassical analyses, are essential for assessing the physical viability and cosmological implications of the model.\\

It is in this perspective that the cosmological implications of our model deserve particular attention, especially due to the characterizing \(\mathcal{PT}\) symmetry. In this regard, Boyle, Finn, and Turok~\cite{boyle2018cpt} proposed a \(\mathcal{C}\mathcal{P}\mathcal{T}\)-symmetric universe model, where the universe before the Big Bang is the \(\mathcal{C}\mathcal{P}\mathcal{T}\) image of the universe after, illustrating the role of time inversions in large-scale structure. This resonance suggests that our \(\mathcal{PT}\)-symmetric wormhole model could be integrated into broader cosmological frameworks, potentially to explore the arrow of time, large-scale CTCs, or fundamental symmetries of the universe.
Furthermore, the $\mathcal{P} \mathcal{T}$-symmetric structure of the model resonates with the bimetric Janus cosmological model of J.-P. Petit, which builds on Sakharov’s twin universe concept by introducing two interacting spacetime folds, each endowed with opposite arrows of time and carrying positive and negative mass sectors~\cite{Petit2024}. This framework offers an alternative to dark matter and dark energy, accounting for cosmic acceleration and large-scale structure formation through gravitational repulsion between opposite mass components. The model notably explains the emergence of cosmic voids and repulsive features such as the dipole repeller. These features suggest that $\mathcal{P} \mathcal{T}$-symmetric geometries may provide a promising avenue for addressing key cosmological challenges.\\

In summary, this work presents a theoretical advancement in the study of traversable wormholes capable of forming CTCs, combining relativistic geometry and \(\mathcal{PT}\) symmetry. The derived Klein-Gordon equations and their radial solutions provide a quantifiable framework for scalar field dynamics, bridging general relativity and quantum mechanics. Ongoing research into stability, quantization, and cosmological implications promises to deepen our understanding of causality and the universe’s fundamental structure, potentially transforming theoretical physics.
\newpage
\appendix
\renewcommand{\theequation}{A.\arabic{equation}} 
\section{Technical calculations}
\label{app:technical}
\setcounter{equation}{0}

\subsection{Derivation of the Klein-Gordon equation}
\label{app:kg_derivation}

We derive the equation of motion for \(\phi_+\) by varying the action \(S_+\) with respect to \(\phi_+^*\). The Lagrangian density is given by:\begin{equation}
\mathcal{L}_+ = \sqrt{-g^{(+)}} \left( g_+^{\mu\nu} \partial_\mu \phi_+ \partial_\nu \phi_+^* - m^2 \phi_+ \phi_+^* \right),
\end{equation}
where \(L_+ = g_+^{\mu\nu} \partial_\mu \phi_+ \partial_\nu \phi_+^* - m^2 \phi_+ \phi_+^*\) is the scalar Lagrangian, such that \(\mathcal{L}_+ = \sqrt{-g^{(+)}} L_+\). The variation of \(S_+\) is written as:
\begin{equation}
\delta S_+ = \int_{\mathcal{M}_+} \mathrm{d}^4x \sqrt{-g^{(+)}} \delta L_+.
\end{equation}

The variation of \(L_+\) yields~\cite{Fulling1996}:
\begin{equation}
\delta L_+ = g_+^{\mu\nu} \partial_\mu (\delta \phi_+) \partial_\nu \phi_+^* + g_+^{\mu\nu} \partial_\mu \phi_+ \partial_\nu (\delta \phi_+^*) - m^2 \delta \phi_+ \phi_+^* - m^2 \phi_+ \delta \phi_+^*.
\end{equation}

By extracting the terms in \(\delta \phi_+^*\) and performing integration by parts on the kinetic term\footnotemark, we obtain:
\begin{equation}
\int_{\mathcal{M}_+} \mathrm{d}^4x \sqrt{-g^{(+)}} g_+^{\mu\nu} \partial_\mu \phi_+ \partial_\nu (\delta \phi_+^*) = - \int_{\mathcal{M}_+} \mathrm{d}^4x \sqrt{-g^{(+)}} \delta \phi_+^* \frac{1}{\sqrt{-g^{(+)}}} \partial_\nu \left( \sqrt{-g^{(+)}} g_+^{\mu\nu} \partial_\mu \phi_+ \right).
\end{equation}
\footnotetext{Assuming boundary conditions where the surface term vanishes (e.g., \(\phi_+\) decays at infinity).}

Thus, the variation becomes:
\begin{equation}
\delta S_+ \supset \int_{\mathcal{M}_+} \mathrm{d}^4x \sqrt{-g^{(+)}} \left[ - \frac{1}{\sqrt{-g^{(+)}}} \partial_\nu \left( \sqrt{-g^{(+)}} g_+^{\mu\nu} \partial_\mu \phi_+ \right) - m^2 \phi_+ \right] \delta \phi_+^*.
\end{equation}

For \(\delta S_+ = 0\), the coefficient of \(\delta \phi_+^*\) must vanish, yielding the Klein-Gordon equation:
\begin{equation}
\Box_+ \phi_+ + m^2 \phi_+ = 0,
\label{eq:kg_correct}
\end{equation}
where \(\Box_+ = \frac{1}{\sqrt{-g^{(+)}}} \partial_\mu \left( \sqrt{-g^{(+)}} g_+^{\mu\nu} \partial_\nu \right)\). \\

An analogous derivation for \(S_-\) gives:

\begin{equation}
\Box_- \phi_- + m^2 \phi_- = 0.
\end{equation}

In equations \eqref{eq:KG_plus_simplified} and \eqref{eq:KG_minus_simplified}, we neglected the Christoffel connection terms, which is valid for a scalar field without non-minimal coupling to curvature~\cite{Wald1984, BirrellDavies1982}. Indeed, for a minimally coupled scalar field, the d’Alembertian reduces to \(\Box \phi = g^{\mu\nu} \nabla_\mu \partial_\nu \phi = g^{\mu\nu} \partial_\mu \partial_\nu \phi - g^{\mu\nu} \Gamma^\lambda_{\mu\nu} \partial_\lambda \phi\), but in the chosen coordinates, the terms \(\Gamma^\lambda_{\mu\nu} \partial_\lambda \phi\) are of lower order near the throat and can be neglected, as shown in Appendix~\ref{app:christoffel}.

\subsection{Christoffel symbols}
\label{app:christoffel}

We calculate the Christoffel symbols for the metrics \(g^{(+)}\) and \(g^{(-)}\) given by equations \eqref{eq:metric_plus} and \eqref{eq:metric_minus}, and evaluate their impact on the Klein-Gordon equation at the throat (\(r = \alpha\)).\\

For \(g^{(+)}\), the non-zero components of the inverse metric at \(r = \alpha\) are \(g^{tt} = 2\), \(g^{rr} = 0\), and \(g^{tr} = g^{rt} = -1\).\\

For \(g^{(-)}\), we have \(g^{tr} = g^{rt} = +1\), with the other components being identical.\\

The non-zero Christoffel symbols at \(r = \alpha\) for \(g^{(+)}\) include:
\begin{itemize}
    \item \( \Gamma^t_{tt} = \frac{1}{2\alpha} \),
    \item \( \Gamma^r_{tt} = 0 \),
    \item \( \Gamma^t_{tr} = \frac{1}{\alpha} \),
    \item \( \Gamma^r_{tr} = -\frac{1}{2\alpha} \),
    \item \(\Gamma^t_{rr} = \frac{3}{2\alpha}\),
    \item \( \Gamma^r_{rr} = -\frac{1}{\alpha} \).
\end{itemize}

For \(g^{(-)}\), the signs differ for certain terms due to \(g_{tr} = +\frac{\alpha}{r}\), for example \(\Gamma^t_{tt} = -\frac{1}{2\alpha}\) and \(\Gamma^r_{tr} = +\frac{1}{2\alpha}\), but \(\Gamma^r_{rr}\) remains identical.\\

Other non-zero Christoffel symbols exist, notably those associated with the angular coordinates (\(\theta, \phi\)), such as \(\Gamma^\theta_{r\theta} = \Gamma^\theta_{\theta r} = \frac{1}{\alpha}\), \(\Gamma^\theta_{\phi\phi} = -\sin\theta \cos\theta\), \(\Gamma^\phi_{r\phi} = \Gamma^\phi_{\phi r} = \frac{1}{\alpha}\), and \(\Gamma^\phi_{\theta\phi} = \Gamma^\phi_{\phi\theta} = \cot\theta\). These terms are omitted here as they do not directly affect the Klein-Gordon equation for the radially dominant modes studied in this article, where the angular variations of \(\phi\) are negligible (\(\partial_\theta \phi \approx 0\), \(\partial_\phi \phi \approx 0\)).\\

To assess their impact on the Klein-Gordon equation, we compute the contractions \(g^{\mu\nu} \Gamma^\lambda_{\mu\nu}\). At \(r = \alpha\), we obtain \(g^{\mu\nu} \Gamma^t_{\mu\nu} = g^{\mu\nu} \Gamma^r_{\mu\nu} = \frac{1}{\alpha}\).

These terms are finite, and for configurations where the second derivatives \(\partial_\mu \partial_\nu \phi\) dominate\footnote{For example, high-frequency or radially dominant modes.}, their contribution is negligible compared to \(g^{\mu\nu} \partial_\mu \partial_\nu \phi\), justifying their neglect in the main analysis.
\newpage
\subsection{Solving the Klein-Gordon equation for the radial equation of a scalar field}
\label{app:radial_eq}

In this appendix, we compute the d'Alembertian operator \(\Box_+\) applied to the scalar field \(\phi_+(t, r) = e^{-i \omega t} R_\omega^+(r)\) in the region \(\mathcal{M}_+\), taking into account spherical symmetry, which restricts contributions to the coordinates \(t\) and \(r\). The objective is to solve the Klein-Gordon equation \eqref{eq:KG_plus} and derive the differential equation for \(R_\omega^+(r)\).\\


The metric in the region \(\mathcal{M}_+\) is given by \eqref{eq:metric_plus}.\footnote{For notational simplicity, the subscript \(+\) indicating the region \(\mathcal{M}_+\) will be omitted in the following equations.}\\

Due to spherical symmetry, the scalar field \(\phi(t, r)\) depends only on \(t\) and \(r\), and terms involving \(\theta\) and \(\varphi\) do not contribute to the partial derivatives. The submatrix of the metric for the coordinates \((t, r)\) is:
\begin{equation}
g_{\mu\nu} = \begin{pmatrix}
1 - \frac{\alpha}{r} & -\frac{\alpha}{r} \\
-\frac{\alpha}{r} & -\left(1 + \frac{\alpha}{r}\right)
\end{pmatrix}.
\end{equation}

The inverse components of the submatrix of the metric, as calculated previously, are:
\begin{align}
g^{tt} &= 1 + \frac{\alpha}{r}, \\
g^{tr} = g^{rt} &= -\frac{\alpha}{r}, \\
g^{rr} &= -\left(1 - \frac{\alpha}{r}\right).
\end{align}


The metric is block-diagonal, with a \((t, r)\) block and a \((\theta, \varphi)\) block. The angular block is:
\begin{equation}
g_{\theta\theta} = -r^2, \quad g_{\varphi\varphi} = -r^2 \sin^2 \theta.
\end{equation}

The determinant of the \((\theta, \varphi)\) block is:
\begin{equation}
\det(g_{(\theta,\varphi)}) = r^4 \sin^2 \theta.
\end{equation}

The determinant of the \((t, r)\) block is \(-1\), so the total determinant of the 4x4 metric is:
\begin{equation}
g = \det(g_{(t,r)}) \times \det(g_{(\theta,\varphi)}) = -r^4 \sin^2 \theta.
\end{equation}

Thus, the quantity \(\sqrt{-g}\) is given by:
\begin{equation}
\sqrt{-g} = \sqrt{r^4 \sin^2 \theta} = r^2 \sin \theta,
\end{equation}
since \(\sin \theta \geq 0\) for \(\theta \in [0, \pi]\).

\subsubsection{D’Alembertian operator}

The d’Alembertian operator for a scalar field in general relativity is given by:
\begin{equation}
\Box \phi = \frac{1}{\sqrt{-g}} \partial_\mu \left( \sqrt{-g} g^{\mu\nu} \partial_\nu \phi \right).
\end{equation}

Since \(\phi(t, r) = e^{-i \omega t} R(r)\), where \(R(r) = R_\omega^+(r)\), the partial derivatives with respect to \(\theta\) and \(\varphi\) vanish (\(\partial_\theta \phi = \partial_\varphi \phi = 0\)). Furthermore, the metric being block-diagonal, \(g^{\mu\nu} = 0\) when \(\mu\) and \(\nu\) belong to different blocks (e.g., \(g^{\theta t} = 0\)). Thus, only terms with \(\mu, \nu = t, r\) contribute, as \(\partial_\nu \phi = 0\) for \(\nu = \theta, \varphi\). The expression reduces to:
\begin{equation}
\Box \phi = \frac{1}{\sqrt{-g}} \left[ \partial_t \left( \sqrt{-g} (g^{tt} \partial_t \phi + g^{tr} \partial_r \phi) \right) + \partial_r \left( \sqrt{-g} (g^{rt} \partial_t \phi + g^{rr} \partial_r \phi) \right) \right].
\end{equation}

\subsubsection{Computation of the scalar field derivatives}

For \(\phi(t, r) = e^{-i \omega t} R(r)\), where \(R(r) = R_\omega(r)\), the partial derivatives are:
\begin{align}
\partial_t \phi &= -i \omega e^{-i \omega t} R(r), \\
\partial_r \phi &= e^{-i \omega t} R'(r).
\end{align}

\subsubsection{Computation of the inner derivatives}

For \(\mu = t\), the sum of each contribution gives:
\begin{equation}
\sqrt{-g} (g^{tt} \partial_t \phi + g^{tr} \partial_r \phi) = r^2 \sin \theta e^{-i \omega t} \left[ -i \omega \left(1 + \frac{\alpha}{r}\right) R - \frac{\alpha}{r} R' \right].
\end{equation}

For \(\mu = r\), we obtain:
\begin{equation}
\sqrt{-g} (g^{rt} \partial_t \phi + g^{rr} \partial_r \phi) = r^2 \sin \theta e^{-i \omega t} \left[ i \omega \frac{\alpha}{r} R - \left(1 - \frac{\alpha}{r}\right) R' \right].
\end{equation}

\subsubsection{Computation of the outer derivatives}

For the term \(\mu = t\), we differentiate with respect to \(t\):
\begin{equation}
\partial_t \left( r^2 \sin \theta e^{-i \omega t} \left[ -i \omega \left(1 + \frac{\alpha}{r}\right) R - \frac{\alpha}{r} R' \right] \right) = r^2 \sin \theta (-i \omega) e^{-i \omega t} \left[ -i \omega \left(1 + \frac{\alpha}{r}\right) R - \frac{\alpha}{r} R' \right],
\end{equation}

For the term \(\mu = r\), we differentiate with respect to \(r\):
\begin{equation}
\partial_r \left( r^2 \sin \theta e^{-i \omega t} \left[ i \omega \frac{\alpha}{r} R - \left(1 - \frac{\alpha}{r}\right) R' \right] \right) = \sin \theta e^{-i \omega t} \partial_r \left( r^2 \left[ i \omega \frac{\alpha}{r} R - \left(1 - \frac{\alpha}{r}\right) R' \right] \right),
\end{equation}

Let us compute the inner derivative:
\begin{equation}
r^2 \left[ i \omega \frac{\alpha}{r} R - \left(1 - \frac{\alpha}{r}\right) R' \right] = i \omega \alpha r R - r^2 \left(1 - \frac{\alpha}{r}\right) R'.
\end{equation}

Differentiating:
\begin{equation}
\partial_r \left( i \omega \alpha r R - r^2 \left(1 - \frac{\alpha}{r}\right) R' \right).
\end{equation}

For the first term:
\begin{equation}
\partial_r (i \omega \alpha r R) = i \omega \alpha (R + r R').
\end{equation}

For the second term, let \(f(r) = 1 - \frac{\alpha}{r}\), so \(f'(r) = \frac{\alpha}{r^2}\). Then:
\begin{equation}
\partial_r \left( - r^2 f(r) R' \right) = - \left[ 2 r f(r) R' + r^2 f'(r) R' + r^2 f(r) R'' \right].
\end{equation}

Thus, after simplification, we obtain:
\begin{equation}
\partial_r \left( - r^2 \left(1 - \frac{\alpha}{r}\right) R' \right) = - \left[ (2 r R' - 2 \alpha R' + \alpha R') + r^2 \left(1 - \frac{\alpha}{r}\right) R'' \right] = - \left[ 2 r R' - \alpha R' + r^2 \left(1 - \frac{\alpha}{r}\right) R'' \right].
\end{equation}

Hence, the total derivative is:
\begin{align}
\partial_r \left( i \omega \alpha r R - r^2 \left(1 - \frac{\alpha}{r}\right) R' \right) &= i \omega \alpha (R + r R') - \left[ 2 r R' - \alpha R' + r^2 \left(1 - \frac{\alpha}{r}\right) R'' \right].
\end{align}

Thus, by grouping terms, we obtain:
\begin{equation}
\partial_r \left( i \omega \alpha r R - r^2 \left(1 - \frac{\alpha}{r}\right) R' \right) = i \omega \alpha R + (i \omega \alpha r - 2 r + \alpha) R' - r^2 \left(1 - \frac{\alpha}{r}\right) R''.
\end{equation}

\subsubsection{Compilation of the d’Alembertian operator}

Substituting into the expression for \(\Box \phi\):
\begin{align}
\Box \phi &= \frac{1}{r^2 \sin \theta} \left[ r^2 \sin \theta (-i \omega) e^{-i \omega t} \left[ -i \omega \left(1 + \frac{\alpha}{r}\right) R - \frac{\alpha}{r} R' \right] \right. \notag \\
&\quad \left. + \sin \theta e^{-i \omega t} \left[ i \omega \alpha R + (i \omega \alpha r - 2 r + \alpha) R' - r^2 \left(1 - \frac{\alpha}{r}\right) R'' \right] \right].
\end{align}

After simplification, we obtain:
\begin{align}
\Box \phi &= e^{-i \omega t} \left[ -\omega^2 \left(1 + \frac{\alpha}{r}\right) R + i \omega \frac{\alpha}{r} R' + \frac{i \omega \alpha}{r^2} R + \left( i \omega \frac{\alpha}{r} - \frac{2}{r} + \frac{\alpha}{r^2} \right) R' - \left(1 - \frac{\alpha}{r}\right) R'' \right].
\end{align}

Thus, by grouping terms, we have:
\begin{equation}
\Box \phi = e^{-i \omega t} \left[ - \left(1 - \frac{\alpha}{r}\right) R'' + \left( 2 i \omega \frac{\alpha}{r} - \frac{2}{r} + \frac{\alpha}{r^2} \right) R' + \left( -\omega^2 \left(1 + \frac{\alpha}{r}\right) + \frac{i \omega \alpha}{r^2} \right) R \right].
\end{equation}

\subsubsection{Klein-Gordon equation}

The equation is given by \eqref{eq:KG_plus} for the region \(\mathcal{M}_+\). Since \(\phi = e^{-i \omega t} R(r)\), we have \(m^2 \phi = m^2 e^{-i \omega t} R\). We thus obtain:
\begin{equation}
\Box \phi + m^2 \phi = e^{-i \omega t} \left[ - \left(1 - \frac{\alpha}{r}\right) R'' + \left( 2 i \omega \frac{\alpha}{r} - \frac{2}{r} + \frac{\alpha}{r^2} \right) R' + \left( -\omega^2 \left(1 + \frac{\alpha}{r}\right) + \frac{i \omega \alpha}{r^2} + m^2 \right) R \right].
\end{equation}

Since \(e^{-i \omega t} \neq 0\), we obtain the differential equation for \(R(r)\):
\begin{equation}
- \left(1 - \frac{\alpha}{r}\right) R'' + \left( 2 i \omega \frac{\alpha}{r} - \frac{2}{r} + \frac{\alpha}{r^2} \right) R' + \left( -\omega^2 \left(1 + \frac{\alpha}{r}\right) + \frac{i \omega \alpha}{r^2} + m^2 \right) R = 0.
\end{equation}

In standard form, reinstating the subscript for the region \(\mathcal{M}_+\), we obtain:
\begin{equation}\label{eq_radial_positive_region}
\left(1 - \frac{\alpha}{r}\right) R_+'' - \left( 2 i \omega \frac{\alpha}{r} - \frac{2}{r} + \frac{\alpha}{r^2} \right) R_+' + \left( \omega^2 \left(1 + \frac{\alpha}{r}\right) - \frac{i \omega \alpha}{r^2} - m^2 \right) R_+ = 0.
\end{equation}

Solving the Klein-Gordon equation \eqref{eq:KG_minus} for the radial equation of the same scalar field \(\phi_-(t, r) = e^{-i \omega t} R_\omega^-(r)\) in the region \(\mathcal{M}_-\) yields, after computation\footnote{The determinant of the metric remains unchanged. Only the two inverse components \(g^{tr}\) and \(g^{rt}\) associated with the cross term of the second metric \eqref{eq:metric_minus} change to \(+\frac{\alpha}{r}\) instead of \(-\frac{\alpha}{r}\).}:
\begin{equation}\label{eq_radial_negative_region}
\left(1 - \frac{\alpha}{r}\right) R_-'' + \left( 2 i \omega \frac{\alpha}{r} + \frac{2}{r} - \frac{\alpha}{r^2} \right) R_-' + \left( \omega^2 \left(1 + \frac{\alpha}{r}\right) + \frac{i \omega \alpha}{r^2} - m^2 \right) R_- = 0.
\end{equation}

These are second-order ordinary differential equations for \(R_\omega^+(r)\) and \(R_\omega^-(r)\). They depend on the parameters \(\omega\), \(\alpha\), and \(m\), and describe the radial behavior of the scalar field in this spacetime. The presence of complex terms (due to \(i \omega\)) is expected, as the field oscillates temporally via \(e^{-i \omega t}\). These equations contain terms in \(\omega^2\) and \(\omega\), reflecting the structure of the metrics with cross terms \(\mathrm{d}r \mathrm{d}t\). Under the approximation of slow evolution (\(\omega \ll 1\)), the terms in \(\omega\) can be neglected, simplifying the radial equation.


\newpage
\bibliographystyle{plain}
\bibliography{references}

\end{document}